\pdfoutput=1 
\documentclass[conference]{IEEEtran}
\IEEEoverridecommandlockouts
\usepackage{cite}
\usepackage{amsmath,amssymb,amsfonts}
\usepackage{algorithmic}
\usepackage{graphicx}
\usepackage{textcomp}
\usepackage{xcolor}


\definecolor{PTB1}{HTML}{4477AA} 
\definecolor{PTB2}{HTML}{EE6677} 
\definecolor{PTB3}{HTML}{228833} 
\definecolor{PTB4}{HTML}{CCBB44} 
\definecolor{PTB6}{HTML}{AA3377} 
\definecolor{PTBR1}{HTML}{2166AC} 
\usepackage{tikz}
\usepackage{booktabs}
\usepackage[version=4]{mhchem}
\DeclareMathOperator{\atan}{atan}
\DeclareMathOperator{\arc2tan}{atan2}
\usetikzlibrary{svg.path}
 
\graphicspath{{figures/}{figures_jul10/}}
\usepackage[capitalize,noabbrev]{cleveref}
    \crefname{figure}{Fig.}{Figs.}
    \crefname{equation}{}{}
    \crefname{section}{Section}{Sections}
    \crefname{table}{Table}{Tables}

\usepackage[group-minimum-digits=4,per-mode=fraction]{siunitx}
\DeclareSIUnit{\ohm}{\ensuremath{\Omega}}
\usepackage[T1]{fontenc}
\usepackage[caption=false,font=footnotesize]{subfig} 
\usepackage{enumitem}
 
\makeatletter
\newcommand{\IEEEleftcaption}[2]{%
  \begingroup
  \long\def\@makecaption##1##2{%
    \@IEEEfigurecaptionsepspace
    \setbox\@tempboxa\hbox{\normalfont\footnotesize {##1.}\nobreakspace\nobreakspace}%
    \parbox[t]{\hsize}{\normalfont\footnotesize\noindent\unhbox\@tempboxa ##2}%
  }%
  \caption{#1}%
  \label{#2}%
  \endgroup
}
\makeatother
 
\usepackage{etoolbox} 
\makeatletter
\patchcmd{\@maketitle}
  {\addvspace{0.5\baselineskip}\egroup}
  {\addvspace{-0.3\baselineskip}\egroup}
  {}
  {}
\makeatother
\definecolor{orcidlogocolor}{HTML}{A6CE39}
\def\orcidsymbolpathouter{svg{M256,128c0,70.7-57.3,128-128,128C57.3,256,0,198.7,0,128C0,57.3,57.3,0,128,0C198.7,0,256,57.3,256,128z}}
\def\orcidsymbolpathinner{svg{M86.3,186.2H70.9V79.1h15.4v48.4V186.2z}
                     svg{M108.9,79.1h41.6c39.6,0,57,28.3,57,53.6c0,27.5-21.5,53.6-56.8,53.6h-41.8V79.1z M124.3,172.4h24.5c34.9,0,42.9-26.5,42.9-39.7c0-21.5-13.7-39.7-43.7-39.7h-23.7V172.4z}
                     svg{M88.7,56.8c0,5.5-4.5,10.1-10.1,10.1c-5.6,0-10.1-4.6-10.1-10.1c0-5.6,4.5-10.1,10.1-10.1C84.2,46.7,88.7,51.3,88.7,56.8z}}
\tikzset{
    orcidlogo/.pic={
        \fill[orcidlogocolor] \orcidsymbolpathouter;
        \fill[white] \orcidsymbolpathinner;
    },
}

\def\orcid#1{\tikz[yscale = -1, scale = 0.035, transform shape, baseline = -7.5pt] \pic{orcidlogo};\,orcid.org/#1}
 
\def\BibTeX{{\rm B\kern-.05em{\sc i\kern-.025em b}\kern-.08em
    T\kern-.1667em\lower.7ex\hbox{E}\kern-.125emX}}
\newcommand{\paperdoi}{}
\usepackage{eso-pic}
\newcommand{\copyrightnotice}{%
    \AddToShipoutPictureBG*{%
        \AtPageLowerLeft{%
            \put(\LenToUnit{0.5\paperwidth},\LenToUnit{0.32in}){%
                \makebox[0pt][c]{\parbox[b]{\dimexpr\paperwidth-1.25in\relax}{\centering\scriptsize
                    Accepted version of a paper presented at the 2026 IEEE Energy Conversion Congress and Expo (ECCE), Vancouver, Canada.%
                    \if\relax\detokenize\expandafter{\paperdoi}\relax\else\ DOI: \paperdoi.\fi\\
                    \copyright\,2026 IEEE. Personal use of this material is permitted. Permission from IEEE must be obtained for all other uses, in any current or future media, including reprinting/republishing this material for advertising or promotional purposes, creating new collective works, for resale or redistribution to servers or lists, or reuse of any copyrighted component of this work in other works.}}}}}}
 
\begin{document}
\copyrightnotice

\title{Improved Single-Input Space Vector for DC-Biased Sinusoidal Current Control in Battery Cyclers
    \thanks{This work was supported by the Natural Sciences and Engineering Research Council of Canada (NSERC) and Magna International Inc. under Alliance Grant ALLRP 595970-2024, and by an NSERC Canada Graduate Research Scholarship - Master's (CGRS-M).
    }
}

\author{
    \IEEEauthorblockN{Manveer Aujla}
    \IEEEauthorblockA{\textit{Electrical and Computer Engineering} \\
        \textit{University of Manitoba}\\
        Winnipeg, Canada \\
        aujlam@myumanitoba.ca\\
        \orcid{0009-0006-9898-140X}}
    \and
    \IEEEauthorblockN{Philip Lewoc}
    \IEEEauthorblockA{
        \textit{Magna International Inc.}\\
        Troy, USA \\
        philip.lewoc@magna.com\\
        \orcid{0009-0008-8679-6164}}
    \and
    \IEEEauthorblockN{Harminderjit S. Toor}
    \IEEEauthorblockA{\textit{Electrical and Computer Engineering} \\
        \textit{University of Windsor}\\
        Windsor, Canada \\
        hstoor@uwindsor.ca\\
        \orcid{0009-0009-3962-7090}}
    \and
    \IEEEauthorblockN{Ahmad Ali}
    \IEEEauthorblockA{\textit{Electrical and Computer Engineering} \\
        \textit{University of Windsor}\\
        Windsor, Canada \\
        ali174@uwindsor.ca\\
        \orcid{0009-0005-4625-6241}}
    \and
    \IEEEauthorblockN{Caniggia Viana}
    \IEEEauthorblockA{\textit{Electrical and Computer Engineering} \\
        \textit{University of Manitoba}\\
        Winnipeg, Canada \\
        caniggia.viana@umanitoba.ca\\
        \orcid{0000-0001-8550-9428}}
    \and
    \IEEEauthorblockN{Narayan C. Kar}
    \IEEEauthorblockA{\textit{Electrical and Computer Engineering} \\
        \textit{University of Windsor}\\
        Windsor, Canada \\
        nkar@uwindsor.ca\\
        \orcid{0000-0002-4082-1888}}
}

\maketitle

\begin{abstract}
    This paper presents improvements to the single-input space vector (SISV) concept for current control in a battery cycler based on a two-phase interleaved buck converter. The cycler is controlled to output a dc-biased sinusoidal current, enabling investigation of second-harmonic ripple effects on lithium-ion cell aging. A modified SISV generation scheme is introduced to track current references containing both dc and sinusoidal components, and an impedance-angle compensation method is proposed to decouple the d and q control axes without disturbing dc tracking. A dq phase-portrait visualization method is also introduced to evaluate transient performance for both SISV and proportional-integral-resonant (PIR) control methods. Simulation and experimental results validate dc-biased sinusoidal current tracking using PIR control and SISV control with different decoupling schemes. The phase-portrait framework allows for straightforward evaluation of transient performance for both SISV and PIR control methods and supports controller tuning to obtain desired trajectories. The results demonstrate the feasibility of SISV-based current control as an alternative to PIR control for battery-cycler applications. The proposed improvements support the development of battery cyclers capable of controlling both dc and sinusoidal current components.
\end{abstract}

\begin{IEEEkeywords}
    Battery cycler, current control, interleaved buck converter, single-input space vector (SISV), rotating reference frame (RRF), control decoupling, proportional-integral-resonant (PIR) control.
\end{IEEEkeywords}

\section{Introduction}

Electric vehicles with modular multilevel converter (MMC)-based drivetrains and battery-integrated converter architectures can distribute energy storage across converter submodules, improve modularity, and enable balancing and fault-tolerant operating modes compared to a single dc-link \cite{Tricoli2017,Tashakor2022DualPort}. However, MMC operation produces second-harmonic power ripple (SHPR) in each submodule, which appears as a ripple current superimposed on the battery-module dc current. The effect of superimposed ac ripple current on lithium-ion (Li-ion) cell aging remains an active research topic. Some prior work suggests that ripple current can contribute to aging \cite{Puranik2018, Kacetl2022Degradation}, while other studies report that harmonics have limited impact on Li-ion cell state of health \cite{Bessman2019}.

Battery degradation depends on operating conditions such as current amplitude, ripple frequency, temperature, and state of charge \cite{Attia2022,Kacetl2022RippleManagement,Kacetl2022AgeingPotential}. Additional experiments across wider current and frequency ranges are needed to quantify the effects of SHPR on MMC battery submodules and support scientific consensus. This need motivates a battery cycler capable of independently controlling a dc current component and a single-tone ac ripple component. Specifically, the control objective is to track a reference of the form
\begin{equation}
    i_o^\star = I_\text{dc} + I_\text{ac}\sin(\omega_0 t).
\end{equation}
While proportional-integral-resonant (PIR) control provides an established stationary-frame solution for dc-biased sinusoidal current tracking \cite{Nazeri2019,Abdulhamid2023,Ghanayem2023}, rotating reference frame (RRF) methods instead represent the sinusoidal component using transformed $d$- and $q$-axis quantities regulated by proportional-integral (PI) compensators \cite{Hackl2016,Zong2012}. The single-input space vector (SISV) method originally proposed in \cite{Hwang2010} provides a framework for applying RRF control concepts to single-phase systems. Still, additional modifications are required when the reference contains both dc and ac components.

This paper improves the SISV method for dc-biased sinusoidal tracking. First, a modified SISV generation scheme is introduced to track references containing both dc and single-tone ac ripple components. Second, an impedance-angle-based method is proposed to decouple the $d$ and $q$ axes without disturbing dc current tracking. Finally, a $dq$ phase portrait is used together with time-domain waveforms to evaluate transient behaviour, current overshoot, and trajectory shape for the different control methods. The PIR and SISV control methods are evaluated in MATLAB/Simulink and validated experimentally on a battery-cycler hardware prototype.

\section{System Analysis and Modelling}
\begin{figure}[htbp]
    \centering
    \includegraphics[width=1\linewidth]{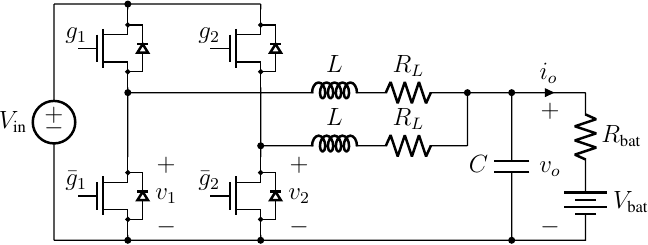}
    \caption{Two-phase interleaved buck converter for battery-cycler application.}
    \label{fig:buck}
\end{figure}

The cycler discussed in the introduction is implemented with a two-phase interleaved buck converter, as shown in \cref{fig:buck}, where the output current $i_o$ is the current entering the battery under test.

\begin{figure}[htbp]
    \centering
    \includegraphics[width=1\linewidth]{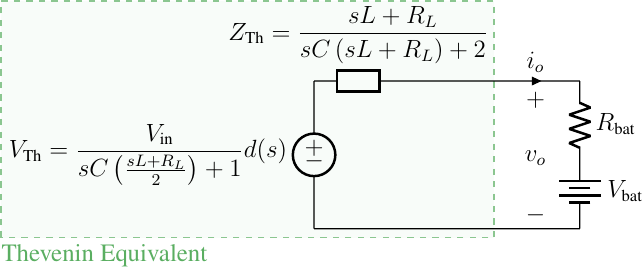}
    \caption{Equivalent circuit for the two-phase interleaved buck converter, taking the Thevenin equivalent from the top terminal of $R_\text{bat}$ and assuming $f_\text{sw} \gg f_0$.}
    \label{fig:thevenin}
\end{figure}

The converter regulates the output current $i_o$ by modulating $v_1$ and $v_2$. Using the same duty cycle $d$ for both phases, the two half-bridge carriers are \SI{180}{\degree} apart to minimize switching ripple. The objective is to regulate the output current to the form $i_{o}^\star=I_\text{dc}+I_\text{ac}\sin\left(2\pi f_0t\right)$, where $f_0$ can be up to \SI{1}{\kilo\hertz} and $I_\text{dc} = I_\text{ac}$ when emulating SHPR. The Thevenin equivalent circuit seen from the top terminal of $R_\text{bat}$ is shown in \cref{fig:thevenin}.

Assuming the switching frequency $f_\text{sw}$ is much greater than the fundamental frequency $f_0$, the averaged converter voltage can be expressed in terms of the steady-state averages of $v_1$ and $v_2$. The average voltage of each half bridge is the same: $\langle v_1 \rangle = \langle v_2 \rangle = dV_\text{in}$. Using voltage division between the series inductor branch and the output capacitor, the Thevenin voltage $V_\text{Th}$ can be expressed as shown in \cref{fig:thevenin} and is proportional to the duty cycle signal $d(s)$. The Thevenin equivalent impedance $Z_\text{Th}$ is also shown in \cref{fig:thevenin}.

Linearizing and neglecting battery-voltage variations gives the duty-cycle-to-output-current transfer function:
\begin{equation}
    \begin{aligned}
        \frac{\tilde{i}_o(s)}{\tilde{d}(s)}
         & =
        \frac{V_\text{in}}
        {\frac{sL + R_L}{2}\left(sCR_\text{bat} + 1\right) + R_\text{bat}} \\[6pt]
         & =
        \left(\frac{2V_\text{in}}{LCR_\text{bat}}\right)
        \frac{1}
        {s^2
            + \left(\frac{R_L}{L} + \frac{1}{CR_\text{bat}}\right)s
            + \frac{R_L + 2R_\text{bat}}{LCR_\text{bat}}}.
    \end{aligned}
    \label{eq:tf}
\end{equation}
A compensator can now be designed using the plant model \cref{eq:tf}. This method can be extended to include additional equivalent series resistances (ESRs) or phases, if needed. If the capacitance is small enough that the circuit is inductive at the reference frequency $f_0$ (i.e., the capacitor is a short circuit to $f_\text{sw}$ and an open circuit to $f_0$), the capacitance can be neglected to obtain a simpler first-order model:
\begin{equation}
    \frac{\tilde{i}_o(s)}{\tilde{d}(s)} = \frac{V_\text{in}}{sL_\text{eq} + R_\text{eq}},
    \label{eq:tf_RL}
\end{equation}
where $L_\text{eq} = \frac{L}{2}$ and $R_\text{eq} = \frac{R_L}{2} + R_\text{bat}$. This simpler model is used to explain the decoupling design in the next section.

\section{Control Design}
This section reviews PIR control and presents the SISV method with different decoupling schemes. The $dq$ framework used for the SISV can also be applied to PIR control to visualize transient performance, as shown in subsequent sections.

\subsection{Proportional-Integral-Resonant (PIR) Control}
To track a sinusoidal reference of a known frequency with a dc component, a PIR compensator is often used:
\begin{equation}
    G_\text{PIR}(s) =
    \underbrace{K_\text{PI}\left(1 + \frac{1}{T_i s}\right)}_{\text{dc control (PI)}}
    +
    \underbrace{K_r\frac{2\omega_0 s}{s^2 + 2\zeta\omega_0 s + \omega_0^2}}_{\text{ac control (R)}},
    \label{eq:PIR}
\end{equation}
where $K_\text{PI}$ is the PI gain, $T_i$ is the integral time constant, $K_r$ is the resonant gain, $\zeta$ is the damping factor, and $\omega_0$ is the reference angular frequency. The PI regulates the dc component and is designed using the plant in \cref{eq:tf_RL}. If $\zeta = 0$, the resonant term has infinite gain at $\omega_0$. In practice, a small but nonzero $\zeta$ is used to improve stability and performance, resulting in finite gain at $\omega_0$ and nonzero steady-state error.

\subsection{Single-Input Space Vector (SISV) Control}
In contrast to PIR control, expressing the reference in anF RRF allows PI compensators to regulate the resulting $dq$ quantities associated with the ac current component. This reduction in control design complexity motivates the use of the SISV method. Zong and Lehn introduce a transformation in \cite{Zong2012} that takes a sinusoidal input signal of a known angular frequency $\omega_0$ and outputs two signals of equal magnitude, one of which is in-phase with the input ($i_\alpha$) and the other lags the input by \SI{90}{\degree} ($i_\beta$). The transfer functions are given by
\begin{subequations}\label{eq:O}
    \begin{align}
        O_{\alpha}(s) & = \frac{i_\alpha^\star(s)}{i_o^\star(s)} = \frac{ks}{s^2 + ks + \omega_0^2}, \label{eq:O_alpha}     \\
        O_{\beta}(s)  & = \frac{i_\beta^\star(s)}{i_o^\star(s)} = \frac{k\omega_0}{s^2 + ks + \omega_0^2},\label{eq:O_beta}
    \end{align}
\end{subequations}
where $k$ sets the filtering and transient response. These blocks are referred to as oscillators because they generate sinusoidal signals at $\omega_0$. Since $O_{\beta}(s)$ has nonzero dc gain, an additional $O_{\alpha}(s)$ block is cascaded with it to reject the dc component. Then the reference current vector and angle can be formed as
\begin{equation}
    \vec{i}_{dq}^\star = i_d^\star + ji_q^\star = \sqrt{{i_\alpha^\star}^2 + {i_\beta^\star}^2} + j0,
    \label{eq:i_dq_ref}
\end{equation}
\begin{equation}
    \theta_0 = \arc2tan\left(i_\beta^\star,\, i_\alpha^\star\right) = \omega_0t.
    \label{eq:theta0}
\end{equation}
The relation $\theta_0 = \omega_0t$ assumes that $i_o^\star$ contains only dc and a single tone at $\omega_0$. Next, a Park transformation is applied to the measured output current using \cref{eq:theta0} as the angle to obtain the output current in the $dq$ frame. PI compensators are then used to generate $d$- and $q$-axis duty cycles, which are brought back to the $\alpha\beta$ frame using an inverse Park transformation with \cref{eq:theta0} again as the angle. The real part, in the $\alpha$-axis, of the ac duty cycle is summed with the dc duty cycle to produce the total duty cycle. The dc and ac duty cycles are saturated between $[-1,1]$, while the total duty cycle is saturated between $[0,1]$.

The dc component is regulated by a separate PI control loop whose bandwidth is much lower than the reference frequency. Since the SISV only operates on the ac component, the dc and ac control loops work independently. If the output SISV tracks the reference SISV in the $dq$ frame, then the output current tracks the reference current in the $\alpha\beta$ frame, which excludes the dc component. Together, the dc and ac control loops work to ensure that the output current tracks the reference current.

\subsection{SISV Control with Conventional Decoupling}
\begin{figure}[htbp]
    \centering
    \includegraphics[width=0.50\linewidth]{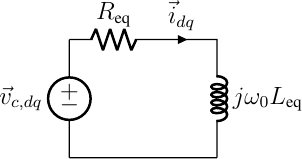}
    \caption{RRF equivalent circuit for simplified model in \cref{eq:tf_RL}.}
    \label{fig:dqCircuit}
\end{figure}

The SISV control-loop transient response can be modified by decoupling the $d$ and $q$ axes. Consider the RRF equivalent circuit in \cref{fig:dqCircuit} based on the simplified model in \cref{eq:tf_RL}, where $\vec{v}_{c,dq}$ is the controlled voltage applied across the impedance and $\vec{i}_{dq}$ is the output current. The output current is
\begin{equation}
    \vec{i}_{dq} = i_{d} + ji_{q} = \frac{v_{c,d} + jv_{c,q}}{R_\text{eq} + j\omega_0 L_\text{eq}}. \label{eq:io_dq}
\end{equation}
Conventional decoupling uses feedforward compensation for the imaginary component of the impedance, which in this simplified model is the inductive voltage drop:
\begin{equation}
    \vec{v}_{L,dq}=j\omega_0 L_\text{eq}\vec{i}_{dq}.
\end{equation}
However, since the oscillators in \cref{eq:O_alpha} and \cref{eq:O_beta} have a transient response, and the output current depends on control performance, $\vec{i}_{dq}$ may contain undesirable transient components during startup. Instead, we use the reference current $\vec{i}_{dq}^\star$ to compute the feedforward compensation.

\subsection{SISV Control with Impedance-Angle-Based Decoupling}
As an alternative to additive decoupling, the multiplicative compensation approach introduced in \cite{Viana2022} can be used. To enforce $\vec{i}_{dq} \rightarrow \vec{i}_{dq}^\star$, a PI compensator can be designed such that
\begin{equation}
    \vec{v}_{c,dq} = K_\text{PI}\left(\vec{e}_{dq} + \frac{1}{T_i}\int{\vec{e}_{dq} \,\mathrm{d}t}\right), \label{eq:PI}
\end{equation}
where $\vec{e}_{dq} = \vec{i}_{dq}^\star - \vec{i}_{dq}$. From \cref{eq:io_dq}, the voltage angle $\angle\vec{v}_{c,dq}$ should converge to the impedance angle if we are to track a purely real current reference, i.e., $i_q^\star=0$. Explicitly,
\begin{equation}
    \angle\vec{v}_{c,dq} \rightarrow \angle Z_\text{eq} = \atan\left(\frac{\omega_0 L_\text{eq}}{R_\text{eq}}\right). \label{eq:Z_angle}
\end{equation}
To request the final voltage angle immediately and mitigate oscillations in the $dq$ frame, we can assign the PI compensator a complex gain,
\begin{equation}
    K_\text{PI} = |K_\text{PI}|e^{j\angle Z_\text{eq}}.
\end{equation}
Equivalently, this rotation can be done by using $\omega_0t + \angle Z_\text{eq}$ as the inverse Park transformation angle. For the more complete model in \cref{eq:tf}, the compensation angle is computed as
\begin{equation}
    \angle Z_\text{eq} = \arc2tan\bigl(\Im\left\{Z_\text{th}\right\},\, \Re\left\{Z_\text{th}\right\} + R_\text{bat}\bigr).
\end{equation}

\begin{figure}[htbp]
    \centering
    \includegraphics[width=1\linewidth]{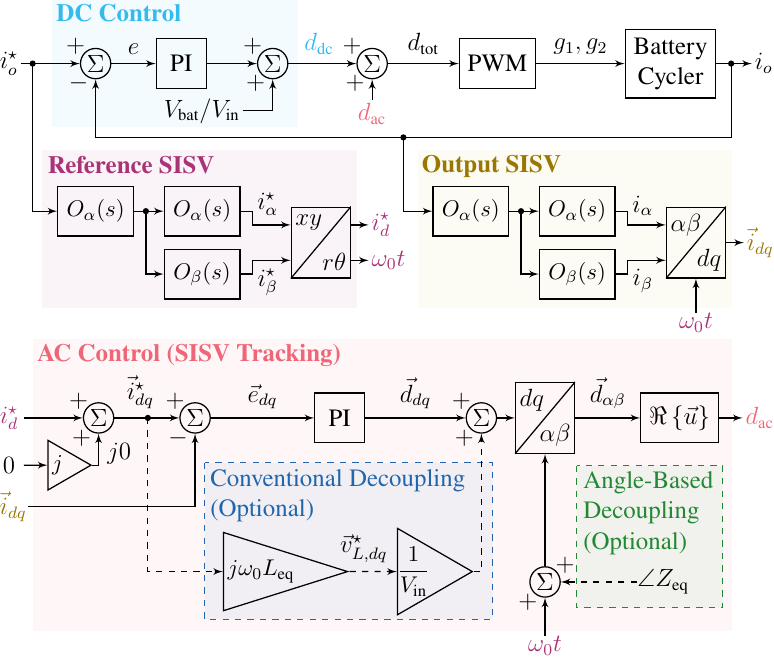}
    \caption{Control block diagram for the improved SISV method for battery-cycler current control. Includes modified SISV generation and optional decoupling using either conventional additive decoupling or impedance-angle-based decoupling.}
    \label{fig:blockDiagram}
\end{figure}

The impedance-angle-based decoupling method has been applied to grid-tied voltage source inverters in \cite{Armita2026}, but this work is the first to apply it using the SISV concept. The complete SISV control block diagram is shown in \cref{fig:blockDiagram}. The term $V_\text{bat}/V_\text{in}$ is added to the dc duty cycle as feedforward compensation for the battery voltage, since this should be the average duty cycle required to maintain the dc current component, neglecting parasitics.

\section{Simulations}
Simulations of a two-phase interleaved buck converter were performed in MATLAB/Simulink to compare the performance of the PIR method, the SISV method without decoupling, the SISV method with conventional decoupling, and the SISV method with angle-based decoupling. A current reference of $i_o^\star = 30\sin(2\pi\cdot 1000t) + 30$\,\si{\ampere} is requested. The simulation parameters, taken directly from the hardware prototype developed for experimental verification, are given in \cref{tb:Params}. Note that the same PI compensator is used for the dc and ac control loops in the SISV method.

\begin{table}[htbp]
    \centering
    \caption{Parameters Used in Simulation and Experimental Verification.}
    \label{tb:Params}
    \renewcommand{\arraystretch}{1.10}
    \small
    \begin{tabular}{
        >{\raggedright\arraybackslash}p{3.5cm} c @{\hspace{0.5cm}} c
        }
        \toprule
        \textbf{Parameter}             & \textbf{Symbol}  & \textbf{Value}                         \\
        \midrule
        \multicolumn{3}{l}{\textbf{System}}                                                        \\
        \addlinespace
        Input Voltage                  & $V_\text{in}$    & \SI{51.68}{\volt}                      \\
        Battery Voltage                & $V_\text{bat}$   & \SI{28}{\volt}                         \\
        Switching Frequency            & $f_\text{sw}$    & \SI{40}{\kilo\hertz}                   \\
        Inductance                     & $L$              & \SI{82}{\micro\henry}                  \\
        Inductor ESR                   & $R_L$            & \SI{10.2}{\milli\ohm}                  \\
        Output Capacitance             & $C$              & \SI{200}{\micro\farad}                 \\
        Capacitor ESR                  & $R_C$            & \SI{16}{\milli\ohm}                    \\
        MOSFET On-Resistance           & $R_\text{ds,on}$ & \SI{10}{\milli\ohm}                    \\
        Battery Resistance             & $R_\text{bat}$   & \SI{102.4}{\milli\ohm}                 \\
        Input Resistance               & $R_\text{in}$    & \SI{10}{\milli\ohm}                    \\
        \midrule
        \multicolumn{3}{l}{\textbf{PIR Controller}}                                                \\
        \addlinespace
        PI Gain                        & $K_\text{PI}$    & \SI{0.0029}{}\,$\text{A}^{\text{--1}}$ \\
        Integral Time Constant         & $T_i$            & \SI{5}{\milli\second}                  \\
        Resonant Gain                  & $K_r$            & \SI{0.0029}{}\,$\text{A}^{\text{--1}}$ \\
        Damping Factor                 & $\zeta$          & 0.01                                   \\
        \midrule
        \multicolumn{3}{l}{\textbf{SISV Controller}}                                               \\
        \addlinespace
        PI Gain                        & $K_\text{PI}$    & \SI{0.0029}{}\,$\text{A}^{\text{--1}}$ \\
        Integral Time Constant         & $T_i$            & \SI{5}{\milli\second}                  \\
        Oscillator Gain in \cref{eq:O} & $k$              & \SI{1000}{}\,$\text{s}^{\text{--1}}$   \\
        \bottomrule
    \end{tabular}
\end{table}

\begin{figure}[htbp]
    \centering
    \includegraphics[width=1\linewidth]{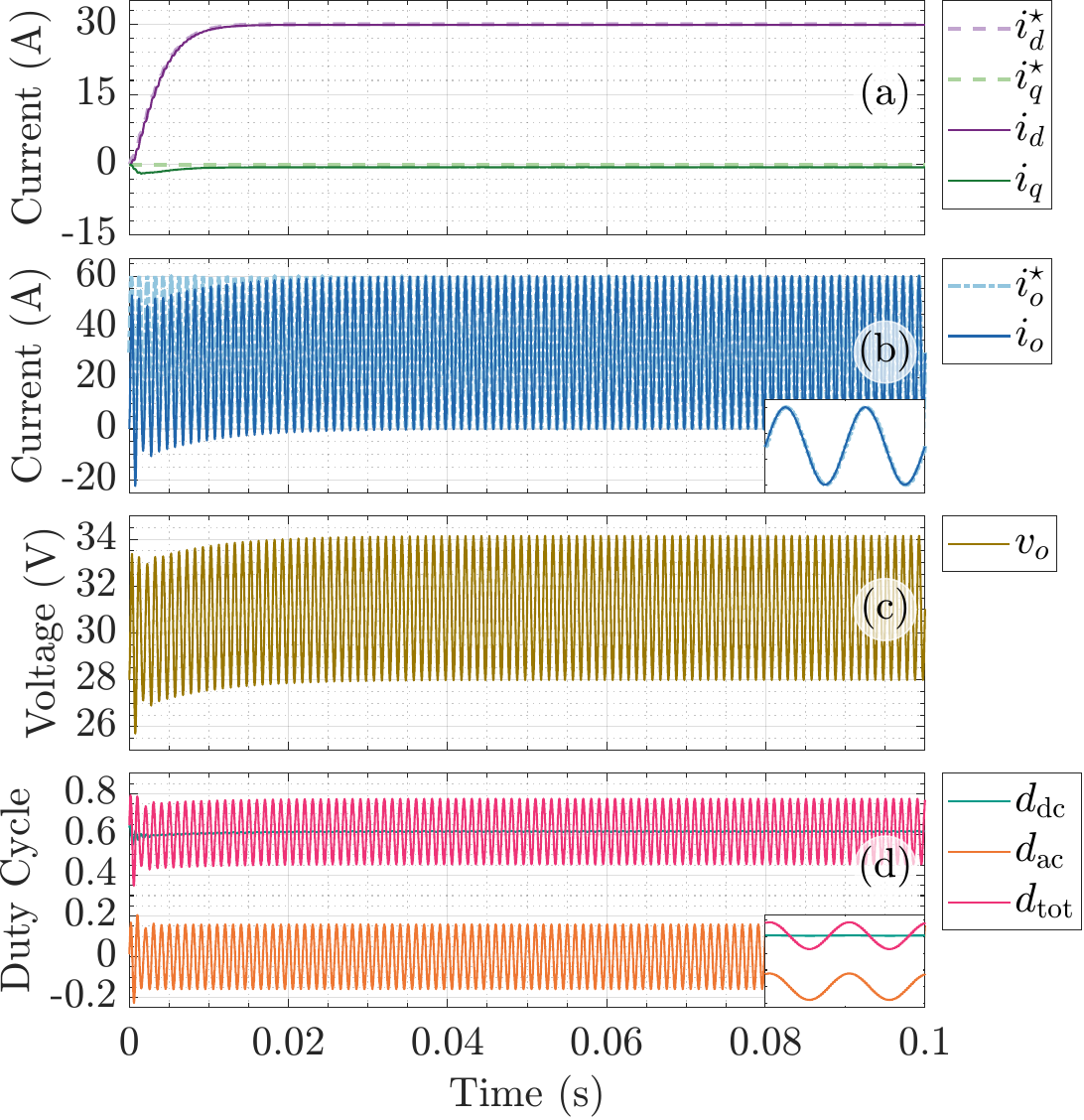}
    \caption{Simulation results for PIR method, $f_0 = \SI{1}{\kilo\hertz}$. (a) Currents in the $dq$ frame. (b) Output current into the battery. (c) Battery terminal voltage. (d) Duty cycles.}
    \label{fig:sim_1kHz_PIR_Plots}
\end{figure}

\begin{figure}[htbp]
    \centering
    \includegraphics[width=1\linewidth]{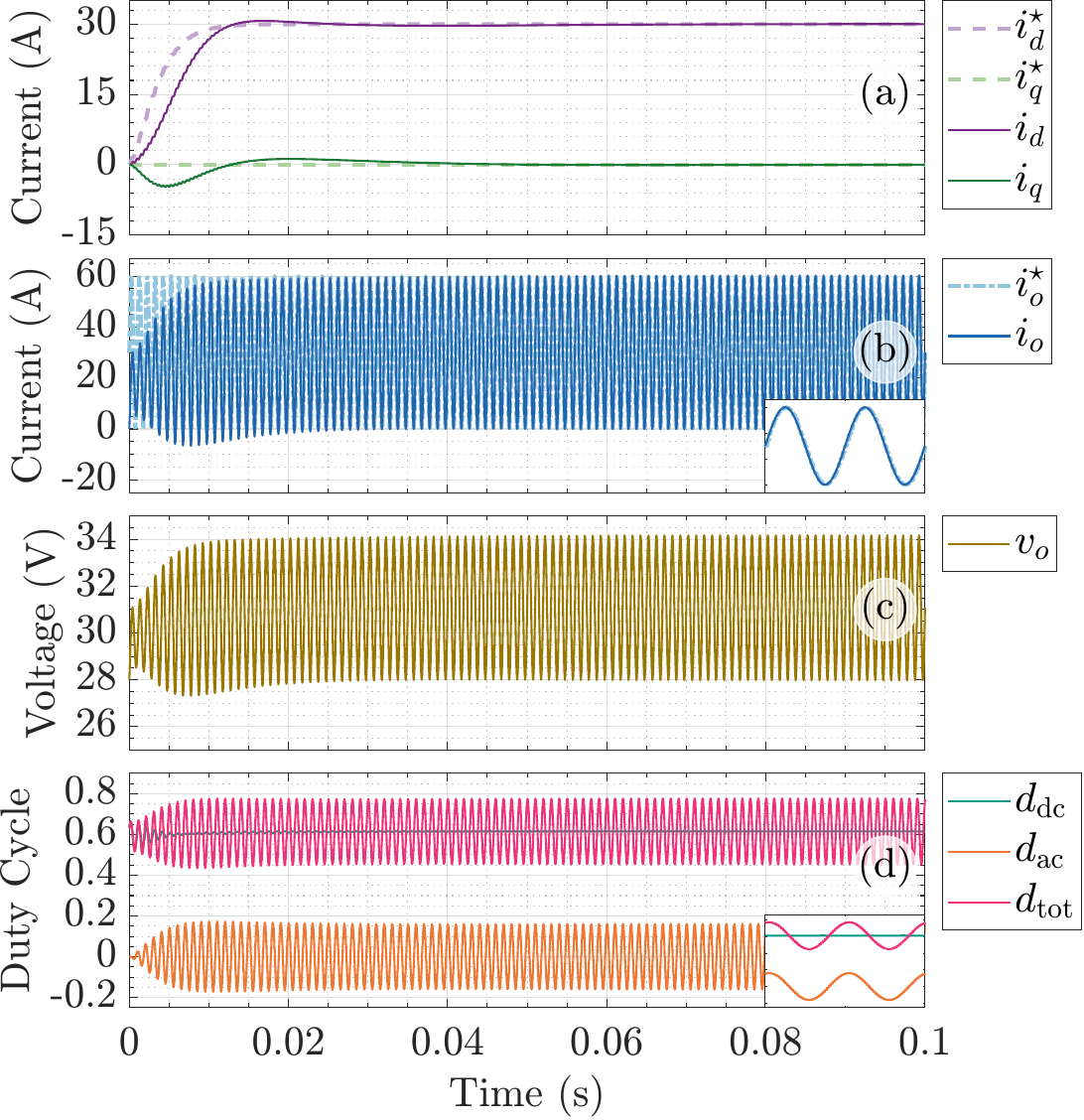}
    \caption{Simulation results for SISV method with conventional decoupling, $f_0 = \SI{1}{\kilo\hertz}$. (a) Currents in the $dq$ frame. (b) Output current into the battery. (c) Battery terminal voltage. (d) Duty cycles.}
    \label{fig:sim_1kHz_dq_DecConv_Plots}
\end{figure}

\begin{figure}[htbp]
    \centering
    \includegraphics[width=1\linewidth]{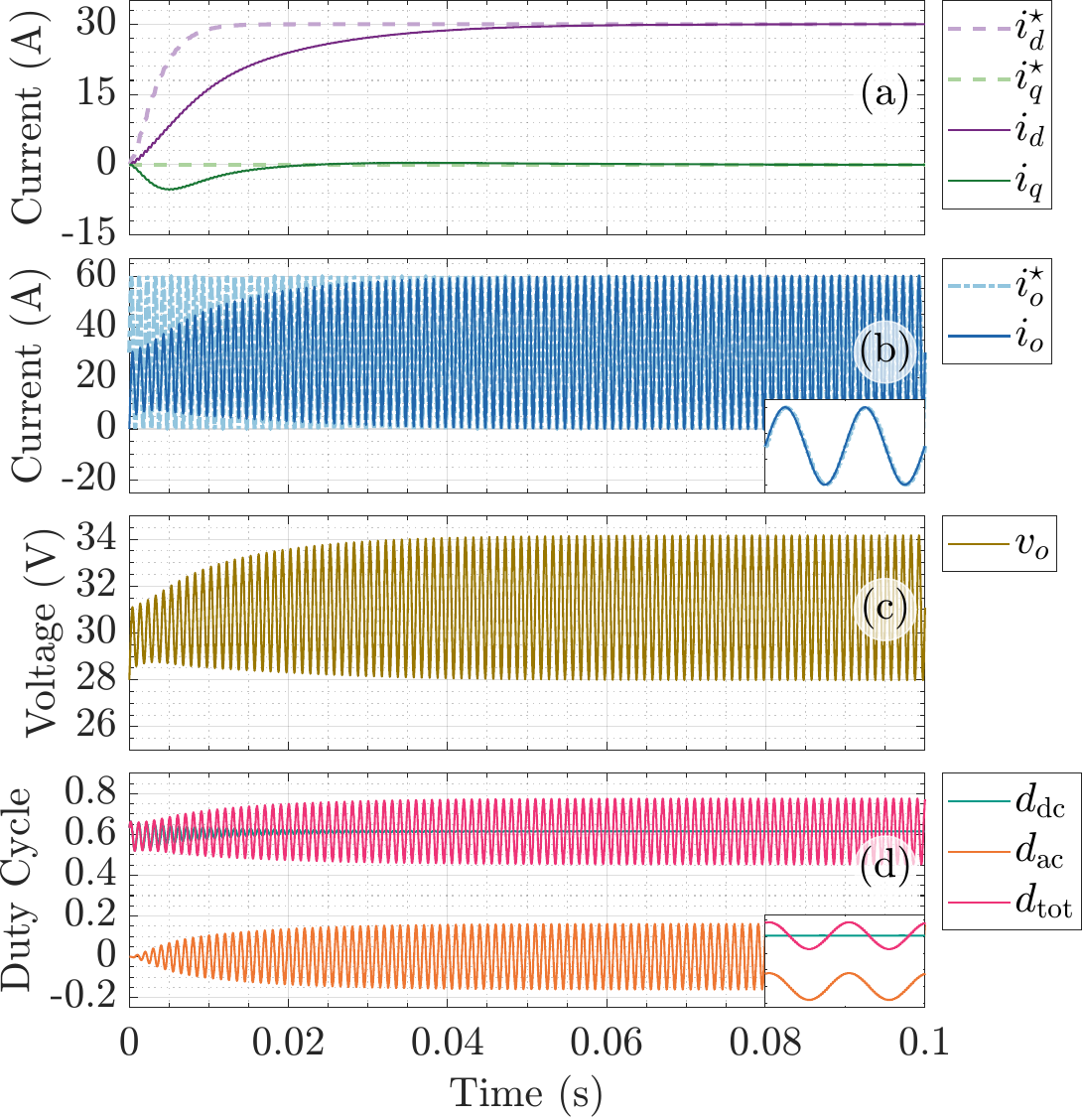}
    \caption{Simulation results for SISV method with angle-based decoupling, $f_0 = \SI{1}{\kilo\hertz}$. (a) Currents in the $dq$ frame. (b) Output current into the battery. (c) Battery terminal voltage. (d) Duty cycles.}
    \label{fig:sim_1kHz_dq_DecAng_Plots}
\end{figure}

The simulations produced waveforms for the $dq$ currents, output current (current into the battery), output voltage (terminal voltage of the battery), and duty cycles (dc, ac, and total). Although the PIR method does not operate in the $dq$ frame, the $dq$ current quantities can still be computed using the same transformations in \cref{fig:blockDiagram}. Simulation results for the PIR method, SISV method with conventional decoupling, and SISV method with angle-based decoupling are shown in \cref{fig:sim_1kHz_PIR_Plots}, \cref{fig:sim_1kHz_dq_DecConv_Plots}, and \cref{fig:sim_1kHz_dq_DecAng_Plots}, respectively. The corresponding plots for the SISV method without decoupling are omitted for brevity.

\begin{figure}[htbp]
    \centering
    \includegraphics[width=0.91\linewidth]{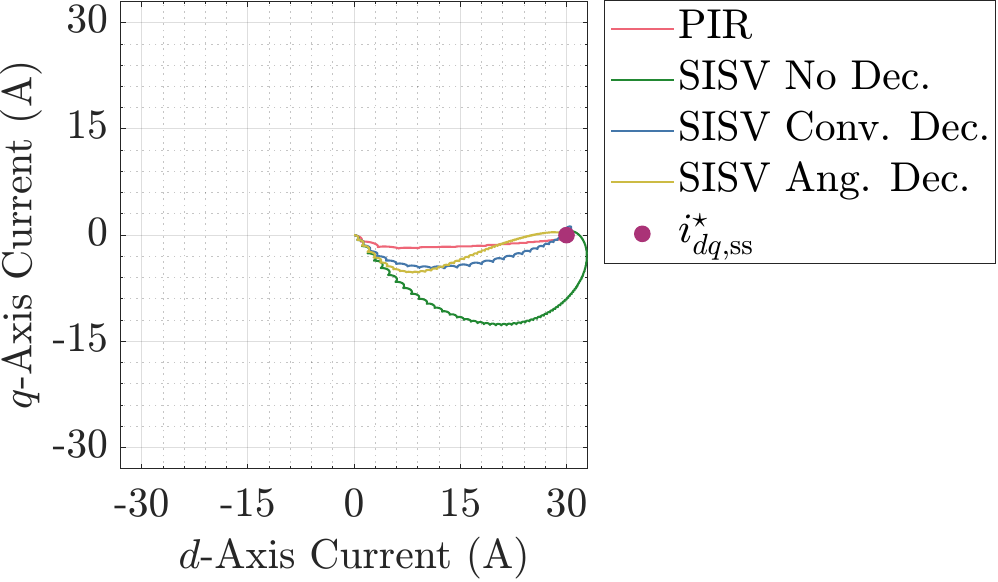}
    \caption{Phase-portrait visualization of $dq$ currents from simulation results for all methods, $f_0 = \SI{1}{\kilo\hertz}$. The reference $dq$ current is $\vec{i}_{dq,\text{ss}}^\star = 30 + j0$\,\si{\ampere}.}
    \label{fig:sim_1kHz_Traj}
\end{figure}

All simulated methods track the reference current. The output voltage demonstrates second-harmonic ripple, and the terminal voltage increases as expected during charging due to the voltage drop across the battery resistance. At steady state, the ac duty cycle is sinusoidal, while the dc duty cycle is constant and begins at the feedforward value of $V_\text{bat}/V_\text{in}$. The total duty cycle is the sum of the dc and ac components.

Further insight into the control performance can be obtained by examining the $dq$-current trajectories toward the steady-state current reference $\vec{i}_{dq,\text{ss}}^\star$ in a phase portrait, as shown in \cref{fig:sim_1kHz_Traj}. The compensated SISV methods follow more direct trajectories toward the reference, while the uncompensated SISV method grows larger in both $d$- and $q$-axis currents before settling. Comparing the two SISV compensation methods, the angle-based decoupling trajectory aligns more closely with the $d$-axis but has slightly more overshoot in the $q$-axis. In the time-domain plots, the angle-based method does not overshoot the output-current reference, whereas the conventional decoupling method does.

In the $dq$ phase portrait, the designed PIR compensator gives the most direct trajectory to the reference. However, it also produces a large negative peak in the output current near \SI{-20}{\ampere}, which is worse than either of the SISV methods. This aggressive behaviour is attributed to the selected PIR gains and can likely be reduced through tuning. This peak is not fully captured in the $dq$ currents because the oscillators require time to build up the $dq$ components. Furthermore, transients much faster than the oscillator frequency are filtered out.

\section{Experimental Results}

An experimental setup was built to validate the proposed battery-cycler current-control methods and is presented in \cref{fig:setup}. The setup consists of an H-bridge PCB connected to an LC filter PCB to form the converter, a Texas Instruments (TI) LaunchPad microcontroller board for digital controls, a lithium iron phosphate (\ce{LiFePO4}, LFP) battery providing the input voltage $V_\text{in}$, and eight series-connected Li-ion battery cells under test in a thermal chamber. The experimental parameters are the same as in the simulation and are listed in \cref{tb:Params}.



\begin{figure}[htbp]
    \centering
    \subfloat[\label{fig:setup1}]{%
        \begin{minipage}{1\linewidth}
            \centering
            \includegraphics[width=1\linewidth]{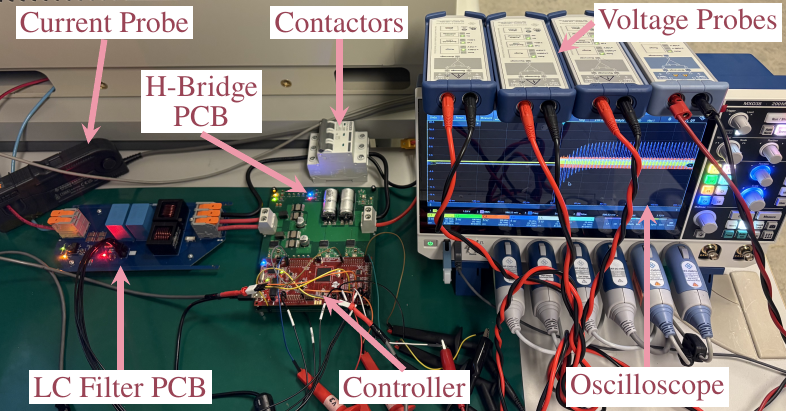}
        \end{minipage}%
    }
    \vspace{0.3em}
    \subfloat[\label{fig:setup2}]{%
        \begin{minipage}{1\linewidth}
            \centering
            \includegraphics[width=0.95\linewidth]{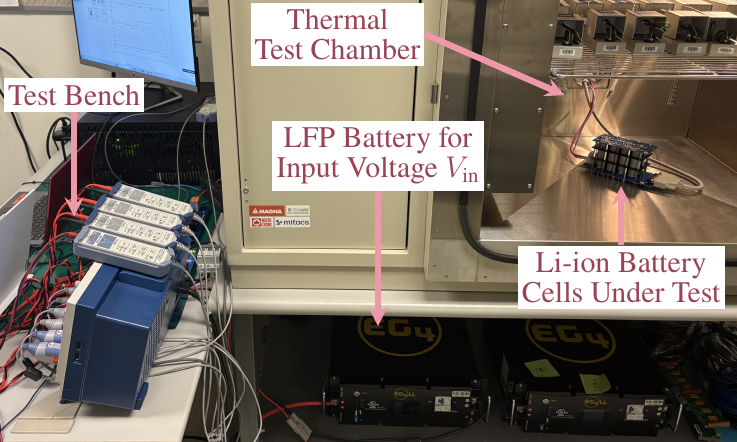}
        \end{minipage}%
    }
    \par
    \IEEEleftcaption{Experimental setup to verify the proposed methods for battery-cycler current control. (a) Test bench with power and control PCBs and measurement equipment. (b) LFP battery providing input voltage $V_\text{in}$ and Li-ion battery cells under test in a thermal chamber.}{fig:setup}
    \vspace{0em}
\end{figure}

The same quantities as in the simulation were measured experimentally, except for the duty cycles. The $dq$ currents were measured from the microcontroller analog outputs using voltage probes, the output current was measured using a current probe, and the output voltage was measured using a voltage probe. The reference current and its $dq$ components were generated in software and overlaid on the plots. A first-order \SI{200}{\hertz} low-pass filter was applied in software to the $dq$ currents to remove high-frequency noise. The output voltage was also filtered in software using a \SI{4}{\kilo\hertz} low-pass filter to remove switching ripple and isolate the SHPR component. The filtered and unfiltered output voltages are both presented.

\begin{figure}[htbp]
    \centering
    \includegraphics[width=1\linewidth]{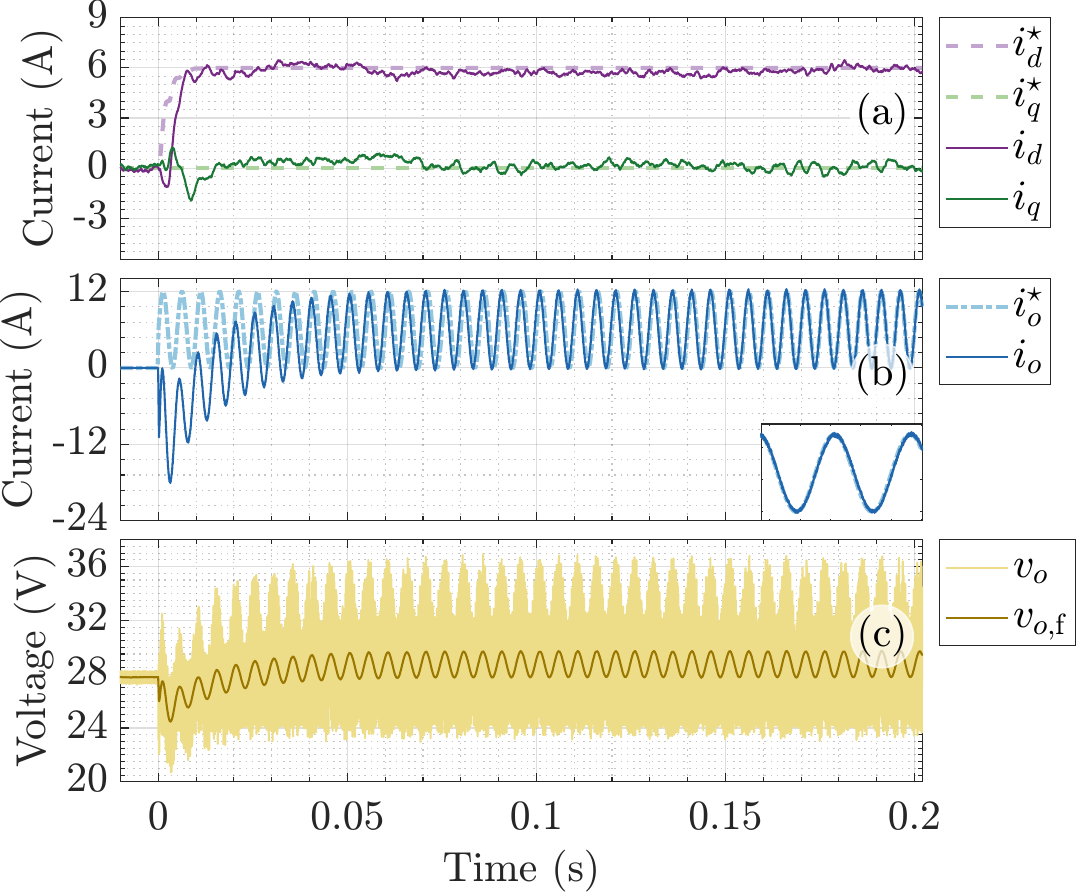}
    \caption{Experimental results for PIR method, $f_0 = \SI{200}{\hertz}$. (a) Currents in the $dq$ frame. (b) Output current into the battery. (c) Battery terminal voltage and software-filtered voltage.}
    \label{fig:exp_PIR_Plots}
\end{figure}

\begin{figure}[htbp]
    \centering
    \includegraphics[width=1\linewidth]{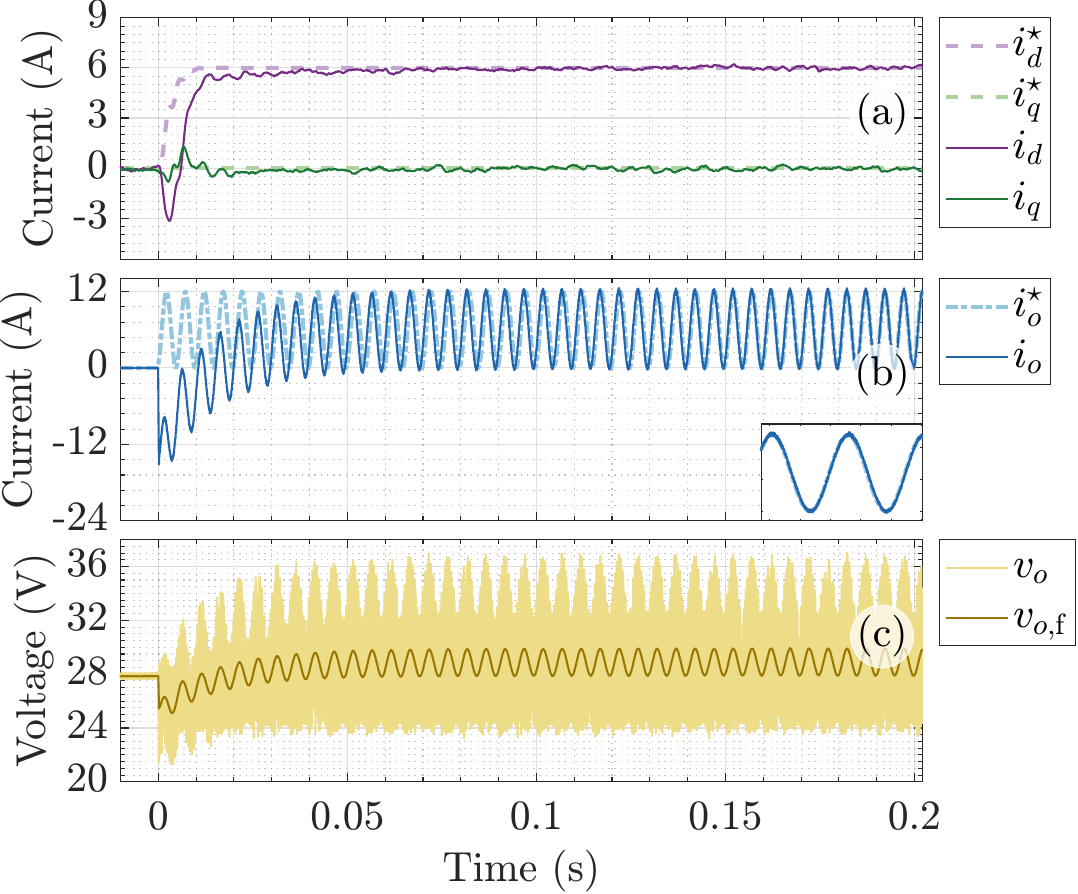}
    \caption{Experimental results for SISV method with conventional decoupling, $f_0 = \SI{200}{\hertz}$. (a) Currents in the $dq$ frame. (b) Output current into the battery. (c) Battery terminal voltage and software-filtered voltage.}
    \label{fig:exp_dq_DecConv_Plots}
\end{figure}

\begin{figure}[htbp]
    \centering
    \includegraphics[width=1\linewidth]{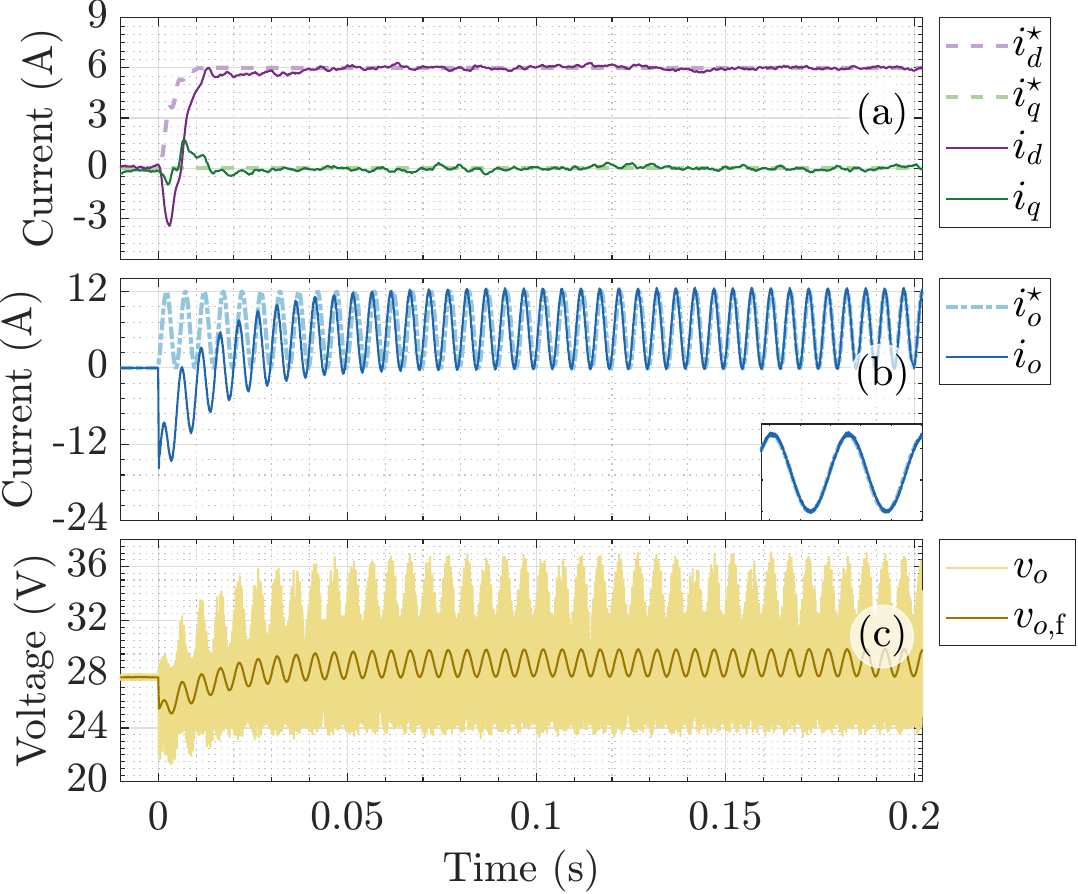}
    \caption{Experimental results for SISV method with angle-based decoupling, $f_0 = \SI{200}{\hertz}$. (a) Currents in the $dq$ frame. (b) Output current into the battery. (c) Battery terminal voltage and software-filtered voltage.}
    \label{fig:exp_dq_DecAng_Plots}
\end{figure}

The requested reference current was $i_o^\star = 6\sin(2\pi\cdot 200t) + 6$\,\si{\ampere}, using smaller currents and a lower frequency than the \SI{1}{\kilo\hertz} simulation case. Extending the implementation to higher frequencies remains future work. Experimental results for the PIR method, SISV method with conventional decoupling, and SISV method with angle-based decoupling are shown in \cref{fig:exp_PIR_Plots}, \cref{fig:exp_dq_DecConv_Plots}, and \cref{fig:exp_dq_DecAng_Plots}, respectively.

\begin{figure}[htbp]
    \centering
    \includegraphics[width=0.92\linewidth]{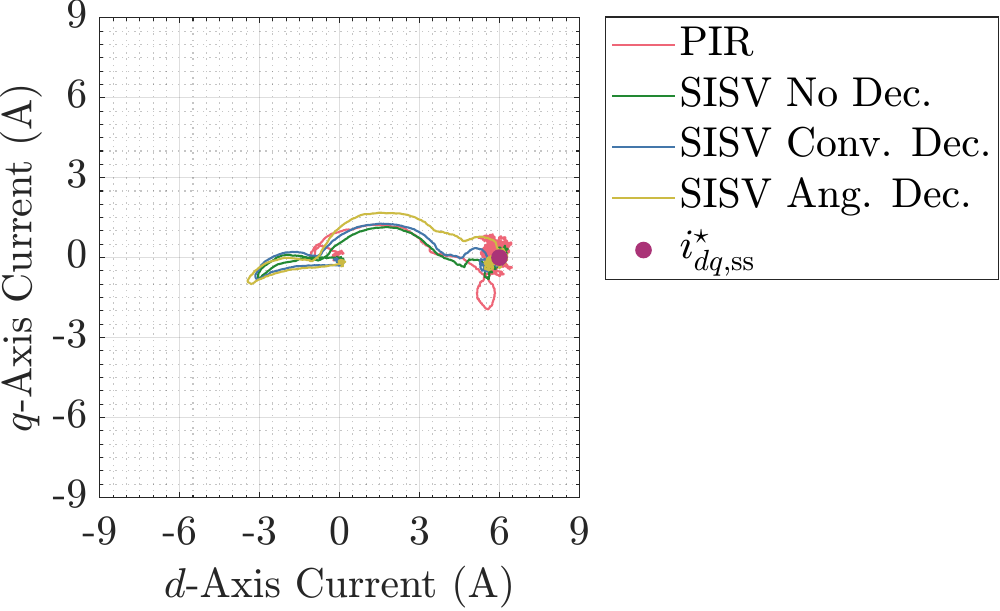}
    \caption{Phase-portrait visualization of $dq$ currents from experimental results for all methods, $f_0 = \SI{200}{\hertz}$. The reference $dq$ current is $\vec{i}_{dq,\text{ss}}^\star = 6 + j0$\,\si{\ampere}.}
    \label{fig:exp_Traj}
\end{figure}

All tested methods track the reference current, but their transient differences are less pronounced at \SI{200}{\hertz} because cross-coupling is proportional to frequency. In all three cases, the output current initially dips below zero, indicating current flow from the battery under test into the cycler input. Subsequent inspection showed that the feedforward duty-cycle term $V_\text{bat}/V_\text{in}$ was computed using $V_\text{bat}=\SI{21.8}{\volt}$ instead of the measured value of $V_\text{bat}=\SI{28}{\volt}$. The resulting initial duty cycle was too low, causing the battery to discharge into the cycler until the dc PI compensator corrected the error.

The $dq$-current trajectories are plotted in \cref{fig:exp_Traj} to evaluate transient performance. Here, the SISV method with angle-based decoupling has the least direct trajectory, while the SISV method without decoupling has the most direct trajectory toward the reference. In the time-domain plots, the PIR method again has the largest negative peak in the output current, while the SISV methods have smaller peaks. Comparing the two decoupled SISV methods, the angle-based method has a larger $q$-axis deviation, which causes its trajectory to be less direct than the conventional decoupling method.

\begin{figure}[htbp]
    \centering
    \includegraphics[width=1\linewidth]{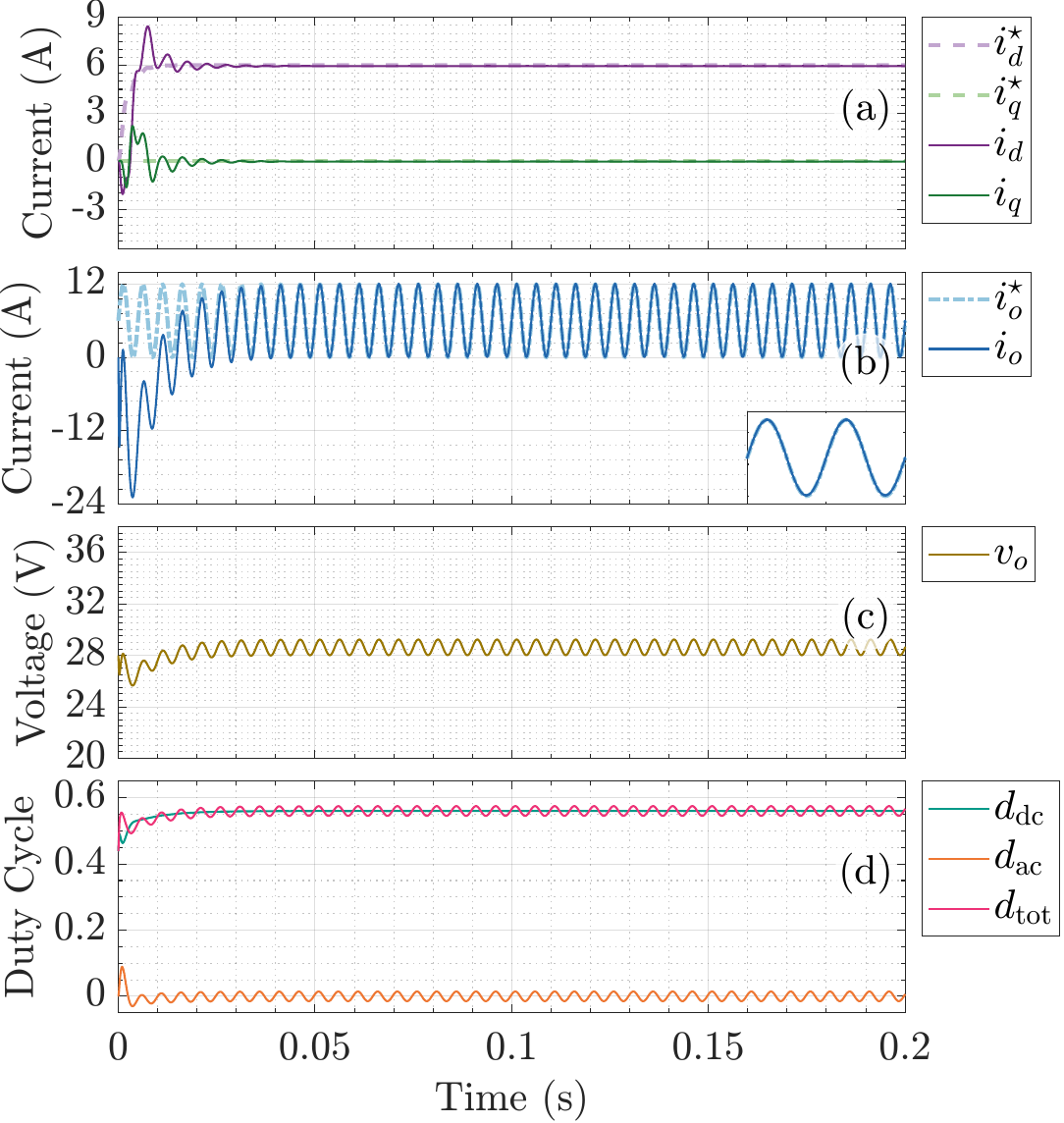}
    \caption{Simulation results for PIR method, $f_0 = \SI{200}{\hertz}$. (a) Currents in the $dq$ frame. (b) Output current into the battery. (c) Battery terminal voltage. (d) Duty cycles.}
    \label{fig:sim_200Hz_PIR_Plots}
\end{figure}

\begin{figure}[htbp]
    \centering
    \includegraphics[width=1\linewidth]{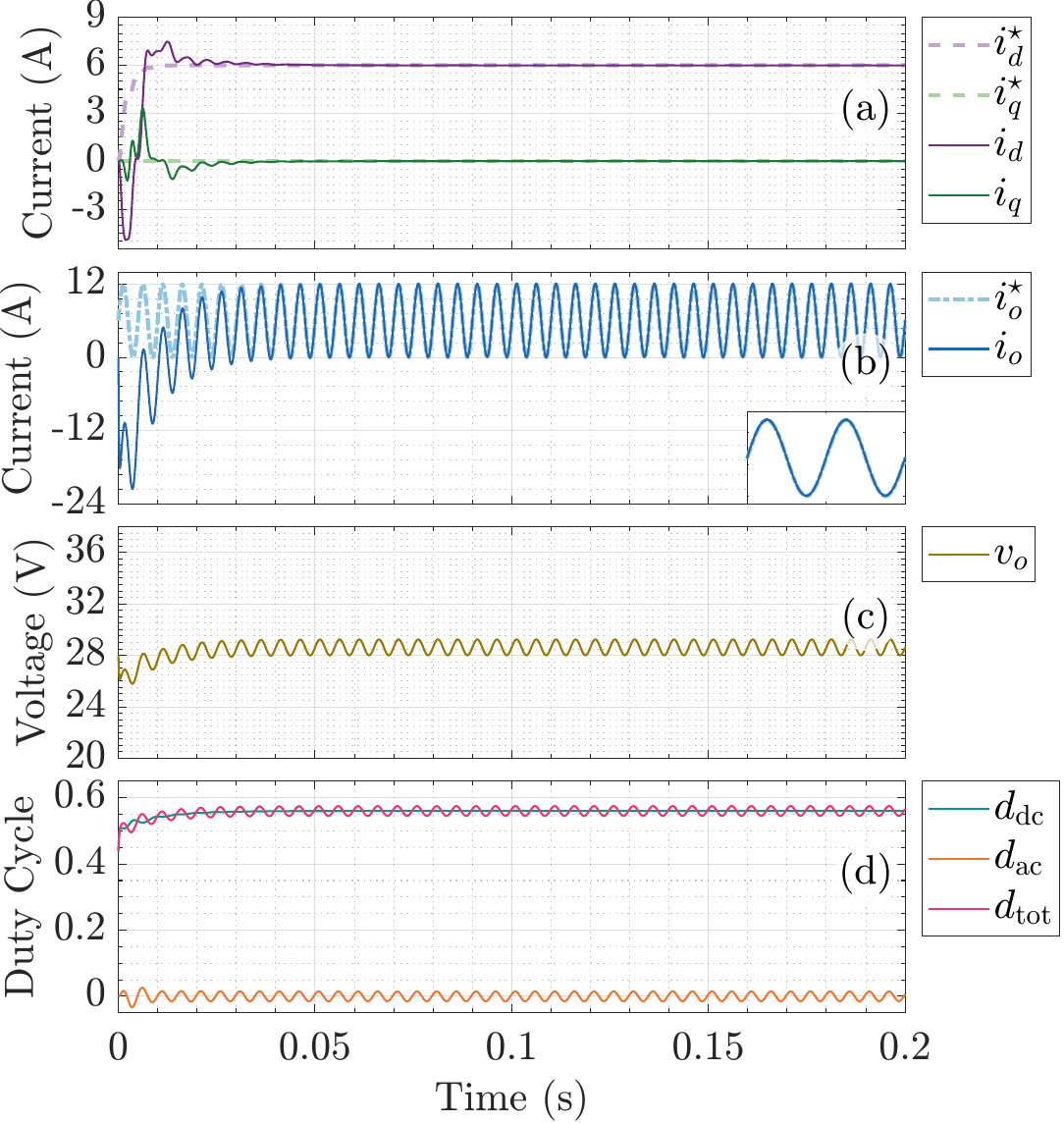}
    \caption{Simulation results for SISV method with conventional decoupling, $f_0 = \SI{200}{\hertz}$. (a) Currents in the $dq$ frame. (b) Output current into the battery. (c) Battery terminal voltage. (d) Duty cycles.}
    \label{fig:sim_200Hz_dq_DecConv_Plots}
\end{figure}

\begin{figure}[htbp]
    \centering
    \includegraphics[width=1\linewidth]{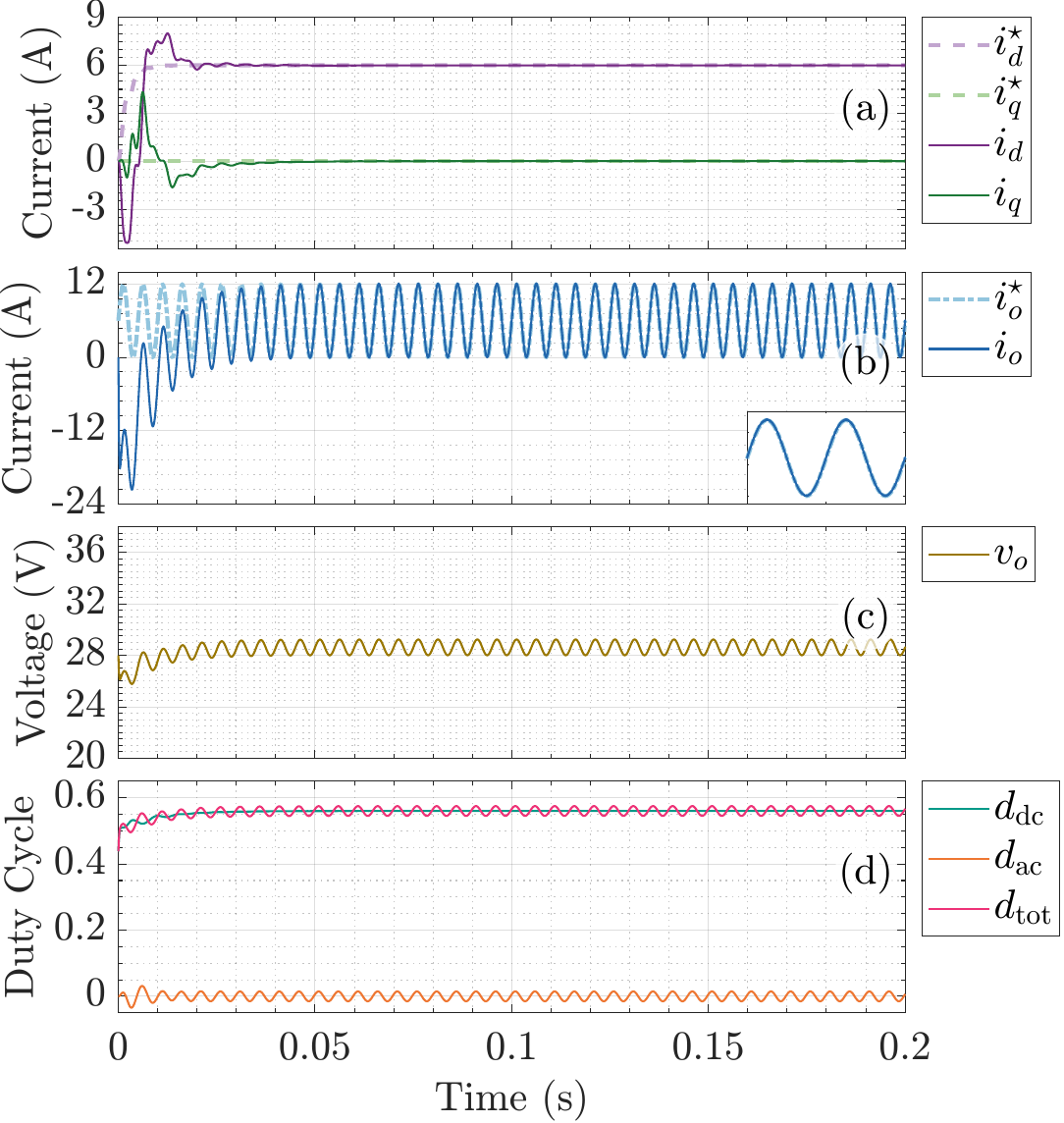}
    \caption{Simulation results for SISV method with angle-based decoupling, $f_0 = \SI{200}{\hertz}$. (a) Currents in the $dq$ frame. (b) Output current into the battery. (c) Battery terminal voltage. (d) Duty cycles.}
    \label{fig:sim_200Hz_dq_DecAng_Plots}
\end{figure}

To better understand the experimental results, the measurements are compared with simulations at the same frequency and with the same feedforward duty-cycle error. The simulation results are shown in \cref{fig:sim_200Hz_PIR_Plots}, \cref{fig:sim_200Hz_dq_DecConv_Plots}, and \cref{fig:sim_200Hz_dq_DecAng_Plots} for the PIR method, SISV method with conventional decoupling, and SISV method with angle-based decoupling, respectively.

\begin{figure}[htbp]
    \centering
    \includegraphics[width=0.92\linewidth]{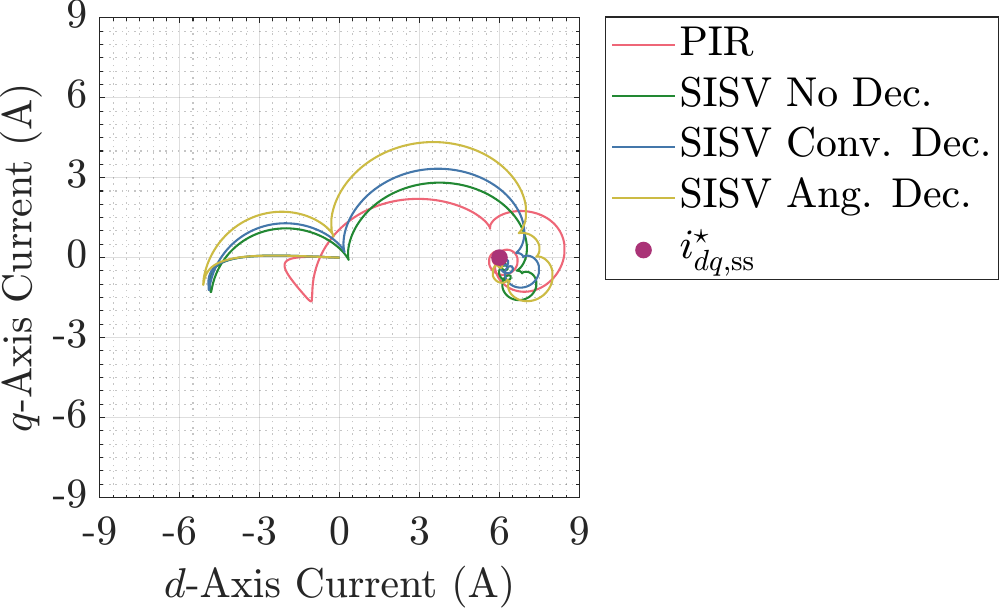}
    \caption{Phase-portrait visualization of $dq$ currents from simulation results for all methods, $f_0 = \SI{200}{\hertz}$. The reference $dq$ current is $\vec{i}_{dq,\text{ss}}^\star = 6 + j0$\,\si{\ampere}.}
    \label{fig:sim_200Hz_Traj}
\end{figure}

The simulation results show that the transient performance remains similar between the three methods at this lower frequency. However, the negative current peaks are much larger in the simulation than in the experimental results, with the PIR method dropping to nearly \SI{-24}{\ampere}. The simulated output-voltage ripple is also smaller than the filtered experimental output-voltage ripple. These discrepancies are likely due to the simulation model not including all parasitic resistances in the experimental setup, such as contact and wiring resistances, and component values that differ from their nominal values (e.g., a larger battery internal resistance).


The $dq$-current trajectories in \cref{fig:sim_200Hz_Traj} show that angle-based decoupling still has the least direct trajectory, while no decoupling has the most direct trajectory. The PIR method also has relatively large oscillations in the $dq$ frame. This is consistent with the experimental results. The reduced benefit of decoupling at \SI{200}{\hertz} is likely due to a mismatch between the circuit model and the plant impedance used for compensation. An inaccurate feedforward term or impedance angle can introduce additional $dq$ oscillations, especially at lower frequencies where the relative resistive, inductive, and capacitive impedance contributions change differently. This likely explains why the uncompensated SISV method gives the most direct trajectory at \SI{200}{\hertz}. Further work is required to improve the impedance model and compensation design for lower-frequency operation.

\section{Conclusion and Future Work}
This paper presents improvements to the SISV method for dc-biased sinusoidal current control in a two-phase interleaved buck battery cycler. A modified SISV generation scheme was introduced to track references with both dc and single-tone ac components. An impedance-angle-based method was proposed to decouple the $d$- and $q$-axis currents, and a $dq$ phase portrait was used with time-domain waveforms to compare transient performance.

The proposed angle-based SISV method was implemented alongside PIR control, SISV without decoupling, and SISV with conventional decoupling. At \SI{1}{\kilo\hertz}, simulation results showed that angle-based decoupling improved the $dq$-current trajectory compared to conventional decoupling by aligning the transient response more closely with the $d$-axis and avoiding output-current overshoot. The PIR method produced the most direct $dq$ trajectory, but also produced the largest negative output-current peak. At \SI{200}{\hertz}, simulation and experimental results showed smaller differences between the control methods because the cross-coupling is smaller at lower frequencies. In this case, decoupling worsened the $q$-axis transient response, likely due to an impedance-model mismatch, while SISV without decoupling produced the most direct trajectory.

The experimental results validate dc-biased sinusoidal current control using both PIR and SISV control methods. Future work will extend the SISV implementation to higher experimental frequencies and improve the impedance model used for decoupling compensation. Further work will also investigate controller tuning using the proposed phase-portrait framework, SISV tracking of multiple frequency components, and the impact of dc and ac duty-cycle saturation on performance. Finally, the experimental cycler setup will be used to evaluate the impact of harmonic current components on battery module state-of-health degradation.

\bibliographystyle{IEEEtran}
\bibliography{IEEEabrv,bibliography}

\end{document}